\documentclass[aps,prb,twocolumn,showpacs]{revtex4}

\usepackage{graphicx}

\begin{document}


\title{Contrasting impurity scattering and pair breaking effects by doping Mn and Zn in
Ba$_{0.5}$K$_{0.5}$Fe$_{2}$As$_{2}$}

\author{Peng Cheng$^1$, Bing Shen$^1$, Jiangping Hu$^{1,2}$ and Hai-Hu
Wen$^1$$^{\star}$}

\affiliation{$^1$National Laboratory for Superconductivity,
Institute of Physics and Beijing National Laboratory for Condensed
Matter Physics, Chinese Academy of Sciences, P.O. Box 603, Beijing
100190, China}

\affiliation{$^2$Department of Physics, Purdue University, West
Lafayette, Indiana 47907, U.S.A.}

\begin{abstract}
Resistivity, Hall effect, magnetoresistance and DC magnetization
were measured in Mn and Zn doped Ba$_{0.5}$K$_{0.5}$Fe$_{2}$As$_{2}$
samples. It is found that the Mn-doping can depress the
superconducting transition temperature drastically with a rate of
$\Delta T_c$/Mn-$1\%$ = -4.2 K, while that by Zn-doping is
negligible. Detailed analysis reveals that the Mn-doping enhances
the residual resistivity ($\rho_0$) significantly, and induces
strong local magnetic moments ($\sim$ 2.58 $\mu_B$) which play as
pair breakers. While the impurity scattering measured by $\rho_0$ in
the Zn-doped samples is much weaker, accompanied by a negligible
pair breaking effect. A possible explanation is that the impurity
scattering by the Zn impurities are mainly small angle scattering
(or small momentum transfer), therefore it cannot break the pairing
induced by the interpocket scattering and thus affect the
superconducting transition temperature weakly.
\end{abstract}

 \maketitle

\section{Introduction}
The discovery of superconductivity above 50 K in iron pnictides has
posed a strong impact in the community of condensed matter
physics.\cite{Kamihara2008} One of the key issues here is about the
superconducting pairing mechanism. Theoretically it was suggested
that the pairing may be established via inter-pocket scattering of
electrons between the hole pockets (around $\Gamma$ point) and
electron pockets (around M point), leading to the pairing manner of
an isotropic gap on each pocket but with opposite signs between them
(the so-called S$^\pm$).\cite{Mazin,Kuroki,LeeDH,LiJX} Meanwhile
other models adopt the S$^\pm$ pairing gap but assume that the
pairing interaction is established via the local magnetic
super-exchange.\cite{HuJP,Graser} Besides, by varying the height of
the pnictogen to the Fe planes, it was argued that the pairing
symmetry may be switched from S$^\pm$ to d-wave
pairing,\cite{AokiPRB} as corroborated by the data in LaFePO where a
nodal gap was inferred from the penetration depth
measurements.\cite{LaFePO,Hicks} Similarly, experimental results
about the pairing symmetry remain highly controversial leaving the
perspectives ranging from S$^{++}$-wave, to S$^\pm$ and to
d-wave.\cite{WYL,Sato,ZhGQ,MuG,Chien,HDing,Hashimoto,LaFePO,Grafe,LuoXG,Prozorov}
Further comprehension to this essential topic is highly desired.

In a superconductor, the disorder induced impurity scattering and
pair breaking strongly depend on the very details of the pairing
gap, therefore it is informative to detect the disorder scattering
effect in the superconducting state. According to the Anderson's
theorem,\cite{AndersonTheorem} in a conventional s-wave
superconductor, nonmagnetic impurities will not lead to apparent
pair-breaking effect. However, a magnetic impurity, owing to the
effect of breaking the time reversal symmetry, can break Cooper
pairs easily. In sharp contrast, in a d-wave superconductor,
nonmagnetic impurities can significantly alter the pairing
interaction and induce a high density of states (DOS) due to the
sign change of the gap on a Fermi surface. This was indeed observed
in cuprate superconductors where Zn-doping induces T$_c$-suppression
as strong as other magnetic disorders, such as Mn and
Ni.\cite{CuprateZnMn} As for the pairing through exchanging the AF
spin fluctuations between different Fermi pockets with the S$^\pm$
pairing, it has been pointed out that nonmagnetic impurities could
severely suppress $T_c$ and the
gap.\cite{Cvetkovic,LeeDHEPL2009,HuJPPRB2009,YuPengWang,Muzikar,Bang,Parker,Onari,NG}
In this paper, we report the doping effect of Mn and Zn to the Fe
sites of superconductor Ba$_{0.5}$K$_{0.5}$Fe$_{2}$As$_{2}$. We
found that the Mn-doping (leading to the magnetic impurities)
depresses T$_c$ drastically, while the impurity scattering and
suppression to T$_c$ by Zn-doping (nonmagnetic scattering centers)
is negligible. The contrasting impurity scattering effects as
revealed by our results need to be reconciled with the theoretical
expectations of the picture of inter-pocket scattering via
exchanging AF spin fluctuations.

\section{Experimental}
The Mn-doped and Zn-doped polycrystalline samples
Ba$_{0.5}$K$_{0.5}$(Fe$_{1-x}$TM$_{x}$)$_{2}$As$_{2}$ (TM = Mn and
Zn) were fabricated by solid state reaction method\cite{Johrendt}.
Powders K$_{3}$As, FeAs, BaAs, ZnAs and MnAs were prepared
previously as precursors. The samples with different doping
concentrations were pressed into pellets under the same pressure,
wrapped in Ta foils and sintered under exactly the same conditions
to eliminate the possible errors in the sample-making process. The
x-ray diffraction (XRD) measurement was performed using an
MXP18A-HF-type diffractometer with Cu-K$_{\alpha}$ radiation. The
analysis of x-ray diffraction data was done by using the softwares
POWDER-X and Fullprof, the obtained results are consistent with each
other. The compositions of the samples were examined by scanning
electron microscopy (SEM, Hitachi S-4200) and the energy dispersive
x-ray analysis (EDX, Oxford-6566). The AC susceptibility
measurements were carried out through an Oxford cryogenic system
Maglab-EXA-12. The resistivity, magnetoresistance and Hall effect
were measured with a Quantum Design instrument physical property
measurement system (PPMS), and the DC magnetization by a Quantum
Design instrument SQUID (MPMS-7).

\begin{figure}
\includegraphics[width=8cm]{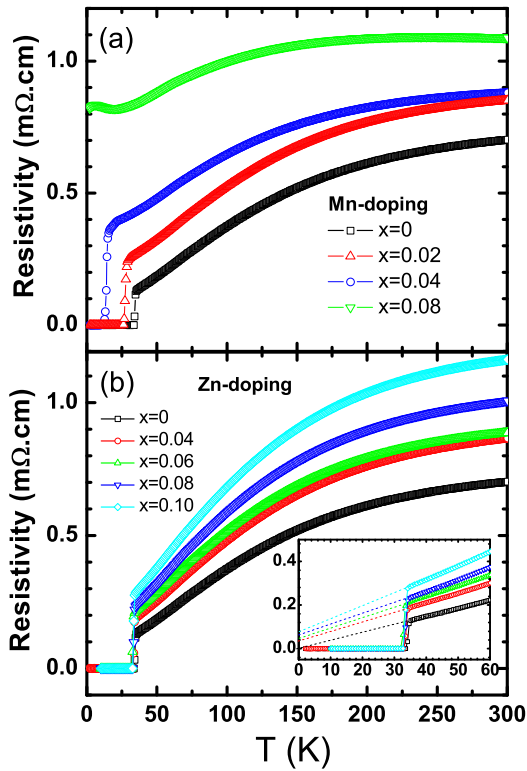}
\caption {(color online) (a) Temperature dependence of resistivity
of the Ba$_{0.5}$K$_{0.5}$(Fe$_{1-x}$Mn$_{x}$)$_{2}$As$_{2}$ samples
under zero field. It is clear that the superconducting transition is
depressed drastically by doping Mn. (b) Temperature dependence of
resistivity of the
Ba$_{0.5}$K$_{0.5}$(Fe$_{1-x}$Zn$_{x}$)$_{2}$As$_{2}$ samples under
zero field. The depression to the superconducting transition by
Zn-doping is negligible. The inset in (b) shows the enlarged part in
the low temperature region. The dashed lines are extrapolation of
the normal state data to T = 0 K. It is remarkable that the undoped
sample has exactly a zero residual resistivity. } \label{fig1}
\end{figure}

\begin{figure}
\includegraphics[width=8cm]{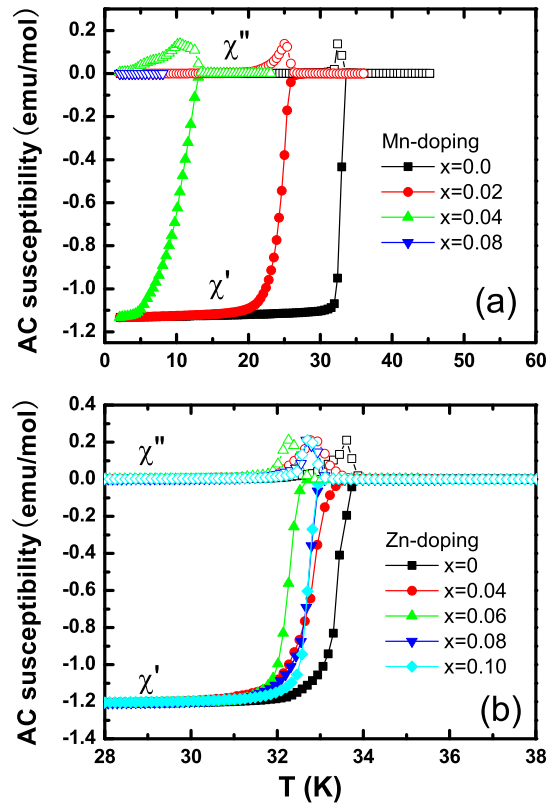}
\caption {(color online) (a) Temperature dependence of AC
susceptibility of the
Ba$_{0.5}$K$_{0.5}$(Fe$_{1-x}$Mn$_{x}$)$_{2}$As$_{2}$ samples
measured with H$_{ac}$ = 0.1 Oe and f=333 Hz. (b) Temperature
dependence of AC susceptibility of the
Ba$_{0.5}$K$_{0.5}$(Fe$_{1-x}$Zn$_{x}$)$_{2}$As$_{2}$ samples
measured with the same conditions as the Mn-doped sample. }
\label{fig2}
\end{figure}

\section{Results}
\subsection{Magnetic and resistive transitions}

In Fig.1(a), we show the temperature dependence of resistivity of
the Mn-doped samples
Ba$_{0.5}$K$_{0.5}$(Fe$_{1-x}$Mn$_{x}$)$_{2}$As$_{2}$. One can see
that the superconducting transition temperature was suppressed
quickly upon the doping of Mn, and the superconductivity vanishes in
the sample with x = 0.08. In contrast, as shown in Fig. 1(b), the
Zn-doped sample does not exhibit a clear change of T$_c$ with the
nominal doping concentration as high as x = 0.10. The Zn-doping
effect in the iron pnictide has been under a heavy debate. It was
found by Li et al.\cite{LiYK} that the Zn-doping to the Fe sites in
LaFeAsO does not change T$_c$ clearly. However, also in the Zn-doped
but high pressure synthesized LaFeAsO samples, Guo et
al.\cite{GuoYF} found that the superconductivity can be destroyed
completely at a very low doping level of 2\%. Actually in our
Zn-doped Ba-122 samples, the slight change of T$_c$ in the samples
with different Zn concentrations may be attributed to the random
scattering of T$_c$ values induced in the synthesizing process. The
temperature dependence of AC susceptibilities for the Mn-doped and
Zn-doped samples are shown in Fig.2(a) and Fig.2(b), respectively.
For Mn-doped ones, the suppression of $T_c$ is remarkable. The
sample with x = 0.08 does not show diamagnetism down to 2 K. This is
consistent with the resistivity data. For Zn-doped ones, however,
there are not much differences in the values of $T_c$ for different
doping levels (x = 0 to 0.1).

\subsection{Structure and composition}

In order to know whether the Zn-impurities are really introduced
into the lattice, we have carried out detailed analysis on the
compositions of the grains in each sample using the EDX analysis.
The undoped sample Ba$_{0.5}$K$_{0.5}$Fe$_{2}$As$_{2}$ exhibits
superconductivity at about $T_c$ = 35 K, combining with the lattice
constants\cite{Johrendt}, we conclude that our sample is slightly
overdoped. The EDX data reveals that the actual doping levels of Mn
is very close to the nominal composition up to 8\%, while Zn doping
has a nonlinear ratio between the really measured composition and
the nominal one: 3.1 $\pm$ 0.3 $\%$ in sample x = 0.04, about 4.5
$\pm$ 0.5 $\%$ in the one x = 0.10. We randomly selected 10 typical
grains and analyzed the compositions on them.Taking the sample with
nominal Zn=10\% as an example, the SEM image and EDX analysis
process were shown in Fig.4 and Table I. This fact indicates that
the Zn impurities have been successfully doped into the lattice,
although the measured composition is lower than the nominal
one.

Above discussion can be corroborated by the characterization of the
XRD data. The XRD patterns for all samples were shown in Fig.3(a)
and Fig.3(b). For Mn-doped ones, one can see that the phase is
rather clean and no impurities could be detected up to x=0.08. The
lattice constants of a-axis and the cell volume increase
monotonically with doping of Mn (as shown in Fig.3(c)), which
indicates that the Mn atoms were successfully introduced into the
lattice. Assuming that the Mn ionic state is "+2", since the ionic
radius of Mn$^{2+}$ (0.8 $\AA$) is bigger than that of Fe$^{2+}$
(0.74 $\AA$), it is understandable that the in-plane lattice
constant expands and the unit cell volume increases about 1.6\%.
While for Zn-doped samples, since the ionic radius for Fe$^{2+}$ and
Zn$^{2+}$ are both 0.74 $\AA$, therefore the distortion is much
smaller, as evidenced by the slight increase of the unit cell volume
(0.34\%). In both systems, it is found that the c-axis lattice
constant does not change obviously compared to the a-axis lattice
constant.
\begin{figure}
\includegraphics[width=8cm]{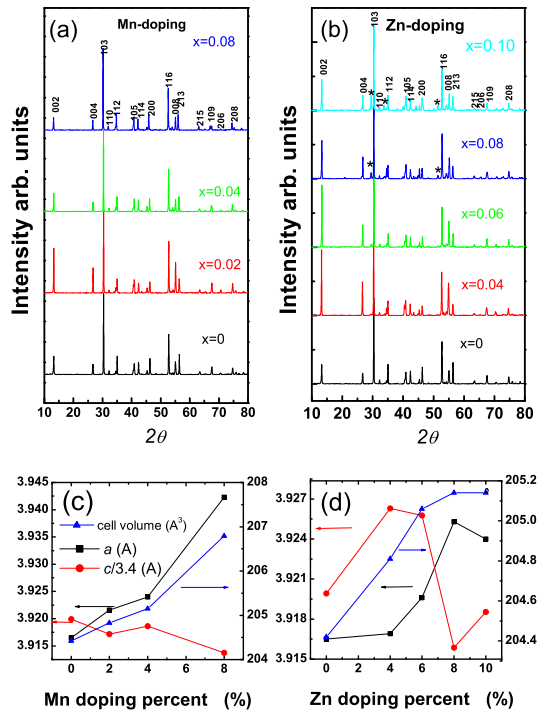}
\caption {(color online) (a) The XRD data for
Ba(Fe$_{1-x}$Mn$_{x}$)$_{2}$As$_{2}$ samples. Up to the doping level
of 8\%, the sample is still quite clean. (b)The XRD data for samples
Ba(Fe$_{1-x}$Zn$_{x}$)$_{2}$As$_{2}$ samples. Slight impurity phase
emerges when the nominal doping composition goes up to 8\%. (c) and
(d) show that the doping dependence of the a-axis and c-axis lattice
constants, as well as the volume of unit cell. One can see that the
a-axis changes much larger in percentage compared with the c-axis
lattice and the volume expansion is mainly dominated by the a-axis
lattice constant.} \label{fig3}
\end{figure}

\begin{figure}
\includegraphics[width=8cm,bb=1 53 238 269]{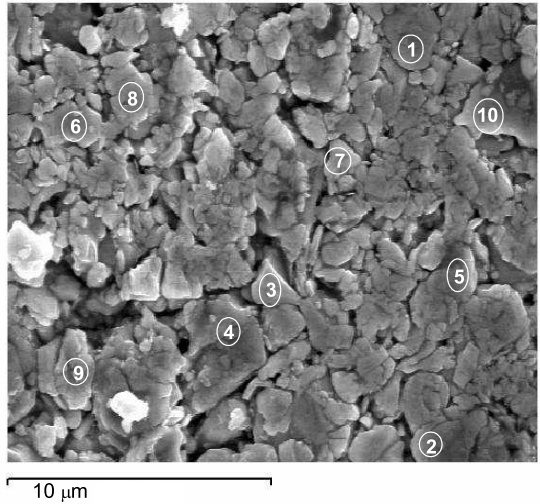}
\caption {(color online) Scanning electron microscopy image of
Zn-doped sample with x=0.10. The numbers in the image correspond to
the typical grains which we chose randomly to carry out the
measurement of energy dispersive x-ray. From the EDX data, the
compositions of Zn in different grains could be obtained and shown
respectively in Table I.} \label{fig4}
\end{figure}

\begin{table}
\caption{the energy dispersive x-ray (EDX) analysis for Zn-doped
sample with x=0.1.}
\begin{tabular}{cccccccc}
\hline \hline
grain & composition & grain & composition \\
\hline
 1   & Zn=4.41\%     & 6       &Zn=4.48\%    &\\
 2   & Zn=4.23\%     & 7       &Zn=3.95\%    &\\
 3   & Zn=4.78\%     & 8       &Zn=4.53\%    &\\
 4   & Zn=5.04\%     & 9       &Zn=4.12\%    &\\
 5   & Zn=4.58\%     & 10      &Zn=4.39\%    &\\

 \hline \hline

\end{tabular}
\label{tab:table1}
\end{table}

\begin{figure}
\includegraphics[width=8cm]{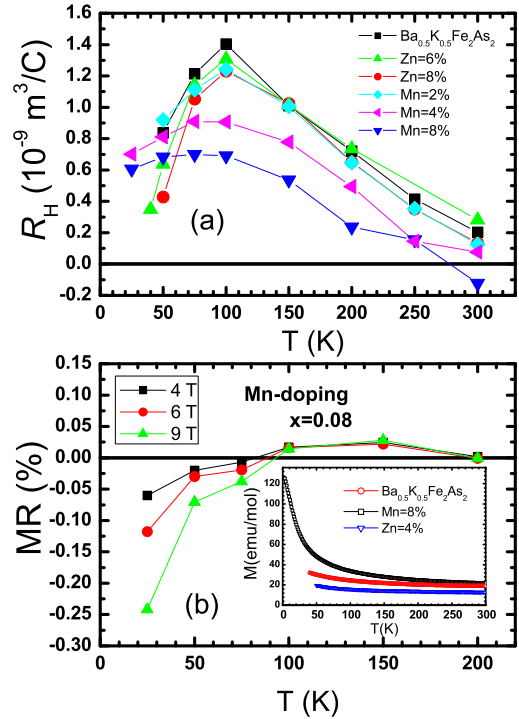}
\caption {(color online) (a) Temperature dependence of Hall
coefficient R$_H$ = $\rho_{xy}$/H measured at 9 T for the undoped,
three Mn-doped and two Zn-doped samples. One can see that by doping
more Mn into the system the positive Hall coefficient R$_H$ becomes
smaller, indicating the doping of holes into the system. While
Zn-doping does not change the Hall coefficient too much. (b) MR of
the Mn-doped samples, a negative MR was observed here. Inset of (b)
shows the temperature dependence of DC magnetization of the undoped,
Mn-doped and Zn-doped sample. The M(T) relation can be described by
the Curie-Weiss law (see text). } \label{fig5}
\end{figure}

\begin{figure}
\includegraphics[width=8cm]{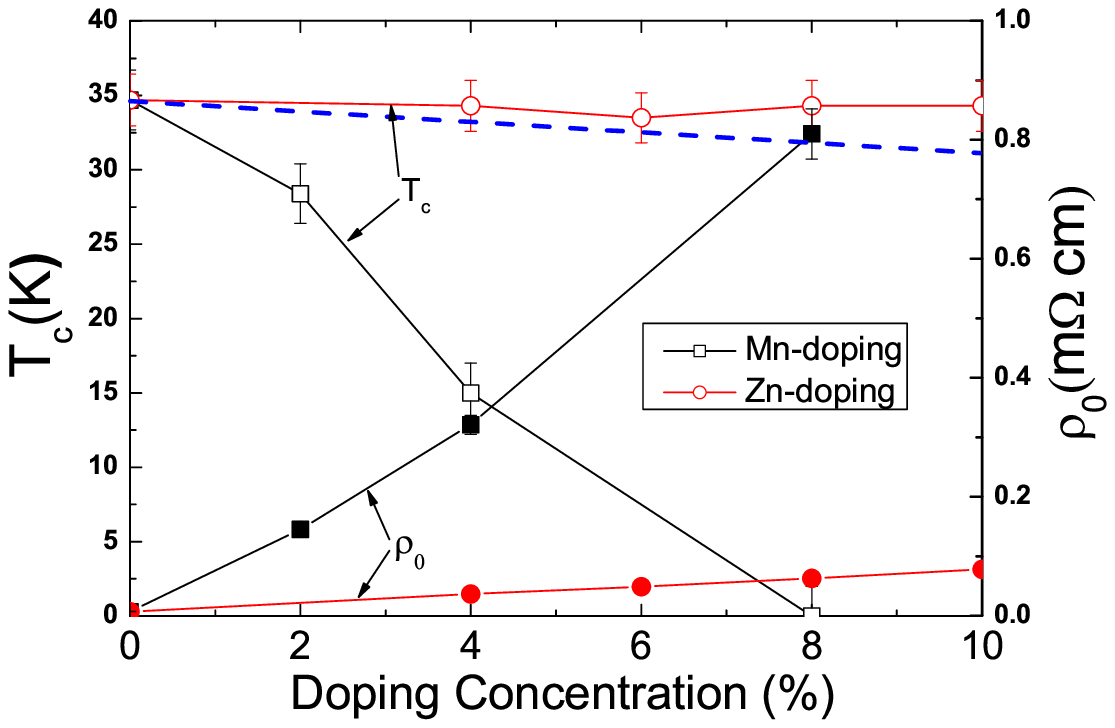}
\caption {(color online) Doping dependence of T$_c$ and $\rho_0$ in
Mn- and Zn-doped samples. The suppression to T$_c$ in Mn-doped
samples is drastic, but that by Zn-doping is negligible. The
enhancement of $\rho_0$ is much weaker in the Zn-doped sample as
compared with the Mn-doped one. The blue dashed line indicates the
T$_c$-drop for the Zn-doped samples if we adopt the simple relation
$k_B\Delta T_c\propto\rho_0$ and taking the same suppression rate of
the Mn-doped system.} \label{fig6}
\end{figure}

\subsection{Hall effect, Magnetoresistance and Magnetizations}
It order to know what has been carried out microscopically through
doping Mn and Zn  in our samples, we measured the Hall coefficient
$R_{H}$, Magnetoresistance (MR) and DC magnetization. As shown in
Fig.5(a), $R_{H}$ is positive in the undoped sample, indicating hole
dominant conduction in Ba$_{0.5}$K$_{0.5}$Fe$_2$As$_2$. As we dope
Mn into the sample and keep increasing the doped content, the R$_H$
reduces systematically. This can be explained by introducing more
holes into the system through Mn-doping, though complexity will be
brought in understanding the Hall data of the multiband
system.\cite{FangL} Actually doping Mn induces similar effect as
doping Cr.\cite{Mandrus} The temperature dependence of R$_H$ is also
shown for the Zn doped sample (x=0.06, 0.08) in Fig. 5(a). One can
see that the data of Zn-doped sample overlap roughly with the
undoped one in a broad temperature region, indicating that the
Zn$^{2+}$ is almost identical to Fe$^{2+}$ in donating electrons.
Fig.5(b) presents the MR of the Mn-doped sample with x = 0.08, a
clear negative MR effect was observed at low temperatures. This
negative MR can be easily understood as due to the enhanced
electron-spin scattering: doping Mn induces more and more magnetic
centers which have stronger magnetic moment compared to that of
Fe$^{2+}$. This argument is supported by the enhanced magnetic
susceptibility in the Mn-doped samples. As shown in the inset of
Fig.5(b), the temperature dependence of DC magnetization of the
Mn-doped sample can be described nicely by the Curie-Weiss law
$\chi=\chi_0+C/(T+T_N$) with $C=\mu_0\mu_J^2/3k_B$, yielding $\mu_J$
= 0.482 $\mu_B$/(Fe) in the undoped sample, 0.645 $\mu_B$/(Fe+Mn) in
the Mn-doped (x=0.08) sample, 0.362 $\mu_B$/(Fe+Zn) in the Zn-doped
(x=0.04) sample. Assuming that each Fe-site has also 0.482 $\mu_B$
in the Mn-doped sample, then each Mn-site contributes about 2.58
$\mu_B$. Zn-doped sample naturally lowers down the paramagnetic
susceptibility, suggesting that nonmagnetic impurities have been
formed. Although it is still under debate whether the AF order is
due to the localized moment or itinerant electrons, it is
qualitatively correct to make an assessment that the increase of the
value of constant "$C$" in the Curie-Weiss law could be attributed
to the enhancement of local magnetic moments upon Mn-doping, this
argument is also consistent with the negative magnetoresistivity
data (Kondo-like scattering).

\section{Discussion}

Finally we summarize the main results in Fig.6. For Mn-doped
samples, as the doping concentration increases, $T_c$ decreases
quickly with a rate of $\Delta T_c$/Mn-$1\%$ = - 4.2 K, while for
the Zn-doped samples the variation of $T_c$ is rather small which
could be termed as negligible. According to the Abrikosov-Gorkov
formula,\cite{AG} if the impurities act as strong pair breakers, the
T$_c$ suppression due to pair breaking is essentially related to the
impurity scattering rate $k_B\Delta T_c\approx
\pi\hbar/8\tau_{imp}\propto\rho_0$, where $\rho_0$ is the residual
resistivity and can be roughly expressed as $m^*/ne^2\tau_{imp}$ in
the single band description with $n$ the charge carrier density,
$m^*$ the effective mass. Therefore we present also the doping
dependence of $\rho_0$ in Fig.6. The enhancement of $\rho_0$ is much
stronger for the Mn-doped samples compared with the Zn-doped ones,
which indicates different impurity scattering effects in the two
sets of samples. For Mn-doped samples, the enhanced average magnetic
moments could act as pair breakers and be responsible for the quick
suppression of superconducting transition temperature, on the other
hand the negative uniaxial chemical pressure effect along the a-axis
(according to the change of lattice constants) may also have
contributions on that. While for Zn-doped samples the residue
resistivity does not increases so much, therefore perhaps the
impurity scattering by the Zn impurities are mainly small angle
scattering (or small momentum transfer), therefore it cannot break
the interpocket scattering pairing. This is understandable since the
Zn$^{2+}$ and Fe$^{2+}$ have very similar ionic sizes.

One may argue that the weak suppression to T$_c$ in the Zn-doped
samples is due to the weak impurity scattering effect (which is
actually also very intriguing), however, even taking this small
change of $\rho_0$ in the Zn-doped samples and assuming the same
suppression rate $\Delta T_c/\Delta\rho_0$ of the Mn-doped system,
we should have a 3.4 K drop of T$_c$ at the doping level of
10\%-Zn/Fe. As marked by the blue dashed line in Fig.6, this is
still outer of the range of the data. This result is consistent with
the impurity scattering effect by doping other transition metals
such as Co and Ni,\cite{Sato2,Hosono} Rh, Ir and Pd to the Fe
sites,\cite{HanFei} where no local strong magnetic moments have been
detected and the superconductivity are rather robust together with
quite strong impurity scattering, as indicated by the large residual
resistivity and small residual resistivity ration (RRR) in these
materials. The weak impurity scattering as well as the pair breaking
effect given by nonmagnetic disorders in iron pnictide may suggest
that these dopants all act as impurity scatters with small momentum
transfer. By changing the doping level, once these scattering vector
in the momentum space is as large as the interpocket vector, we
shall see a strong pair breaking effect, although it is
non-magnetic. It is very worthwhile to check this interesting
scenario.

Before concluding the paper, we should mention another interesting
discovery in our data. It is found that
$[\rho(T,x)-\rho(0,x)]/[\rho(300K,x)-\rho(0,x)]$ is doping
independent in all temperature regions and all data with different
doping levels can be nicely scaled (see Fig.7). This indicates that,
in the single band approach, the
$[m^*(x)/\tau(x,T)n(x,T)e^2]/[m^*(0)/\tau(0,T)n(0,T)e^2]$ is
temperature independent. Regarding the similar temperature
dependence of all the curves, we would assume that
$1/\tau(x)\approx1/\tau(0)$, therefore the temperature independent
scaling is due to the electron-boson coupling strength $\lambda$ in
terms of the effective mass relation $M^*(x)=(1+\lambda M^*(0))$.
Suppose that adding Zn has two effects: (1) adding impurity
scattering and (2) changing the coupling strength with spin
fluctuations (bosons here). The latter assumption is pretty logical
in view of the strong effect of Zn on the magnetic properties of the
parent compound. If this is true, Zn doping may enhance the
superconducting pairing, but unfortunately it is offset by its
pair-breaking effect as an impurity. It is very worthwhile to check
whether this scaling holds also in other systems.

\begin{figure}
\includegraphics[width=8cm]{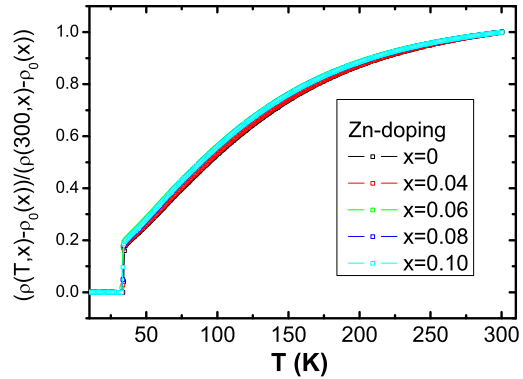}
\caption {(color online) Temperature dependence of
$[\rho(T,x)-\rho(0,x)]/[\rho(300K,x)-\rho(0,x)]$ for Zn-doped
samples with different doping levels, where $\rho(T,x)$ represents
the resistivity of the sample at temperature "$T$" and with doping
level "$x$".} \label{fig7}
\end{figure}

\section{Concluding remarks}
We fabricated and measured the resistivity, Hall effect,
magnetoresistance and DC magnetization in Mn and Zn doped
Ba$_{0.5}$K$_{0.5}$Fe$_{2}$As$_{2}$ samples. The Mn-doping enhances
the residual resistivity and suppresses the superconducting
transition temperature drastically. Further analysis indicates that
the Mn-doping induces strong local magnetic moments ($\sim$ 2.58
$\mu_B$) which play as pair breakers. The impurity scattering
measured by the residual resistivity $\rho_0$ in the Zn-doped
samples is very weak, accompanied by a negligible suppression to
T$_c$. A possible explanation is that the impurity scattering by the
Zn impurities are mainly small angle scattering (or small momentum
transfer), therefore it cannot break the pairing given by the
interpocket scattering. Finally we found a scaling of the normal
state resistivity: $[\rho(T,x)-\rho(0,x)]/[\rho(300K,x)-\rho(0,x)]$.
This is explained as the enhancement of the coupling strength
between electrons and the spin fluctuations induced by the Zn
doping.

\begin{acknowledgments}
We appreciate the useful discussions with I. I. Mazin, D.-H. Lee and
G.-M. Zhang. This work is supported by the NSF of China, the
Ministry of Science and Technology of China (973 projects:
2006CB601000, 2006CB921107, 2006CB921802), and Chinese Academy of
Sciences within the knowledge innovation programm.
\end{acknowledgments}

$^{\star}$ hhwen@aphy.iphy.ac.cn


\begin{thebibliography}{99}

\bibitem{Kamihara2008} Y. Kamihara, T. Watanabe, M. Hirano, and H. Hosono, J. Am. Chem. Soc. {\bf130}, 3296 (2008).
\bibitem{Mazin} I. I. Mazin, D. J. Singh, M. D. Johannes, and M. H. Du, Phys. Rev. Lett. {\bf101}, 057003 (2008).
\bibitem{Kuroki} K. Kuroki, S. Onari, R. Arita, H. Usui, Y. Tanaka, H. Kontani, and H. Aoki, Phys. Rev. Lett. {\bf101}, 087004 (2008).
\bibitem{LeeDH} F. Wang, H. Zhai, Y. Ran, A. Vishwanath, and D. H. Lee, Phys. Rev. Lett. {\bf102}, 047005 (2009).
\bibitem{LiJX}Z. J. Yao, J. X. Li, and Z. D. Wang, New J. Phys. {\bf11}, 025009 (2009).
\bibitem{HuJP}K. Seo, B. A. Bernevig, and J. P. Hu, Phys. Rev. Lett. {\bf101}, 206404(2008).
\bibitem{Graser} S. Graser, T. A. Maier, P. J. Hirschfeld, and D. J. Scalapino
New J. Phys. {\bf11}, 025016 (2009).
\bibitem{AokiPRB}K. Kuroki, H. Usui, S. Onari, R. Arita, and H. Aoki, Phys. Rev. B {\bf79}, 224511 (2009).
\bibitem{LaFePO} J. D. Fletcher, A. Serafin, L. Malone, J. Analytis, J-H Chu, A. S. Erickson, I. R. Fisher, A. Carrington. Phys. Rev. Lett. {\bf102}, 147001 (2009).
\bibitem{Hicks} C. W. Hicks, T. M. Lippman, M. E. Huber, J. G. Analytis, J. H. Chu, A. S. Erickson, I. R. Fisher, and K. A. Moler, Phys. Rev. Lett. {\bf103}, 127003(2009).
\bibitem{WYL} Y. L. Wang, L. Shan, L. Fang, P. Cheng, C. Ren, and H. H. Wen, Supercond. Sci. Technol. {\bf22}, 015018 (2009).
\bibitem{Sato} T. Sato, S. Souma, K. Nakayama, K. Terashima, K. Sugawara, T. Takahashi, Y. Kamihara, M. Hirano, and H. Hosono, J. Phys. Soc. Jpn. {\bf77}, 063708 (2008).
\bibitem{ZhGQ} S. Kawasaki, K. Shimada, G. F. Chen, J. L. Luo, N. L. Wang, and G. Q. Zheng, Phys. Rev. B {\bf78}, 220506(R) (2008).
\bibitem{MuG} G. Mu, X. Y. Zhu, L. Fang, L. Shan, C. Ren, and H. H. Wen, Chin. Phys. Lett. {\bf25}, 2221 (2008). G. Mu, H. Q. Luo, Z. S. Wang, L. Shan, C. Ren, and H. H. Wen, Phys. Rev. B {\bf79}, 174501 (2009).
\bibitem{Chien} T. Y. Chen, Z. Tesanovic, R. H. Liu, X. H. Chen, C. L. Chien, Nature (London) {\bf453}, 1224 (2008).
\bibitem{HDing} H. Ding, P. Richard, K. Nakayama, T. Sugawara, T. Arakane, Y. Sekiba, A. Takayama, S. Souma, T. Sato, T.
Takahashi, Z. Wang, X. Dai, Z. Fang, G. F. Chen, J. L. Luo, and N.
L. Wang, Europhys. Lett. {\bf83}, 47001 (2008).
\bibitem{Hashimoto}K. Hashimoto, T. Shibauchi, T. Kato, K. Ikada, R. Okazaki, H. Shishido, M. Ishikado, H. Kito, A. Lyo, H. Eisaki, S. Shamoto, and Y. Matsuda, Phys. Rev. Lett.
{\bf102}, 017002 (2009).
\bibitem{Grafe}H.-J. Grafe, D. Paar, G. Lang, N. J. Curro, G. Behr, J. Werner, J. H. Borrero, C. Hess, N. Leps, R. Klingeler, and B. Buchner, Phys. Rev. Lett.
{\bf101}, 047003(2009).
\bibitem{LuoXG}X. G. Luo, M. A. Tanatar, J. P. Reid, H. Shakeripour, N. Doiron-Leyraud, N. Ni, S. L. Budko, P. C. Canfield, H. Q. Luo, Z. S. Wang, H. H. Wen, R. Prozorov, and L. Taillefer, Phys. Rev. B {\bf 80}, 140503(R) (2009).
\bibitem{Prozorov}R. T. Gordon, N. Ni, C. Martin, M. A. Tanatar, M. D. Vannette, H. Kim, G. D. Samolyuk, J. Schmalian, S. Nandi, A. Kreyssig, A. I. Goldman, J. Q. Yan, S. L. Budko, P. C. Canfield, and R. Prozorov, Phys. Rev. Lett. {\bf102}, 127004 (2009)
\bibitem{AndersonTheorem} P. W. Anderson, J. Phys. Chem. Solids {\bf11}, 26 (1959).
\bibitem{CuprateZnMn}G. Xiao, M. Z. Cieplak, J. Q. Xiao, and C. L. Chien, Phys. Rev. B{\bf42}, 8752 (1990).
\bibitem{Cvetkovic}V. Cvetkovic and Z. Tesanovic, Europhys. Lett. {\bf85}, 37002 (2009).
\bibitem{LeeDHEPL2009}F. Wang, H. Zhai and D. H. Lee, Europhys. Lett. {\bf85}, 37005
(2009).
\bibitem{HuJPPRB2009} Y. Y. Zhang, C. Fang, X. T. Zhou, K. J. Seo, W. F. Tsai, B. A. Bernevig, and J. P. Hu,
Phys. Rev. B {\bf 80}, 094528(2009).
\bibitem{YuPengWang}J. Li, and Y. P. Wang, Europhys. Lett. {\bf88}, 17009 (2009).
\bibitem{Muzikar} G. Preosti, H. Kim, and P. Muzikar, Phys. Rev. B {\bf50}, 1259 (1994).
\bibitem{Bang} Y. Bang, H. Choi, and H. Won, Phys. Rev. B {\bf79}, 054529 (2009).
\bibitem{Parker} D. Parker O. V. Dolgov, M. M. Korshunov, A. A. Golubov, and I. I. Mazin, Phys. Rev. B {\bf78}, 134524 (2008).
\bibitem{Onari} S. Onari, and H. Kontani, Phys. Rev. Lett. {\bf103}, 177001
(2009).
\bibitem{NG} T. K. Ng, and Y. Avishai, arXiv: 0906.2442.
\bibitem{Johrendt}M. Rotter, M. Tegel, and D. Johrendt, Phys. Rev. Lett. {\bf101}, 107006 (2008).
\bibitem{LiYK}Y. K. Li, X. Lin, Q. Tao, C. Wang, T. Zhou, L. J. Li, Q. B. Wang, M. He, G. H. Cao, and Z. A. Xu, New J. Phys. {\bf11}, 053008 (2009).
\bibitem{GuoYF}Y. F. Guo, Y. G. Shi, S. Yu, A. A. Belik, Y. Matsushita, M. Tanaka, Y. Katsuya, K. Kobayashi, I. Nowik, I. Felner, V. P. S. Awana, K. Yamaura, E. T. Muromachi, Condmat:/Arxiv:0911.2975
\bibitem{FangL} L. Fang, H. Q. Luo, P. Cheng, Z. S. Wang, Y. Jia, G. Mu, B. Shen, I. I. Mazin, L. Shan, C. Ren, H. H. Wen, Phys. Rev. B {\bf80}, 140508 (R) (2008).
\bibitem{Mandrus}A. S. Sefat, D. J. Singh, L. H. VanBebber, Y. Mozharivskyj, M. A. McGuire, R. Y. Jin, B. C. Sales, V. Keppens, and D. Mandrus, Phys. Rev. B {\bf 79}, 224524 (2009).
\bibitem{AG}A. A. Abrikosov and L. P. Gor'kov, Soviet Phys. JETP {\bf12}, 1243 (1961).
\bibitem{Sato2}M. Sato, Y. Kobayashi, S. C. Lee, H. Takahashi, E. Satomi, Y. Miura, J. Phys. Soc. Jpn. {\bf79}, 014710 (2010).
\bibitem{Hosono}S. Martsuishi, Y. Inoue, T. Nomura, Y. Kamihara, M. Hirano and H. Hosono, New J. Phys. {\bf11}, 025012 (2009).
\bibitem{HanFei}F. Han, X. Y. Zhu, P. Cheng, G. Mu, Y. Jia, L. Fang, Y. L. Wang, H. Q. Luo, B. Zeng, B. Shen, L. Shan, C. Ren, and H. H. Wen, Phys. Rev. B {\bf80}, 024506 (2008).



\end{thebibliography}
\end{document}